\documentclass[conference]{IEEEtran}
\IEEEoverridecommandlockouts
\usepackage{cite}
\usepackage{amsmath,amssymb,amsfonts}
\usepackage{algorithmic}
\usepackage{graphicx}
\usepackage{textcomp}
\usepackage{xcolor}

\usepackage{array}
\usepackage{booktabs}
\usepackage{tabularx}
\usepackage{bigstrut}
\usepackage{multirow,diagbox,tabularx,blindtext}
\usepackage{hyperref}

\usepackage[ruled,linesnumbered]{algorithm2e}

\newcommand{\PP}[1]{
\vspace{2px}
\noindent{\bf \IfEndWith{#1}{.}{#1}{#1.}}
}

\def\BibTeX{{\rm B\kern-.05em{\sc i\kern-.025em b}\kern-.08em
    T\kern-.1667em\lower.7ex\hbox{E}\kern-.125emX}}

\begin{document}
\title{FedRecAttack: Model Poisoning Attack to Federated Recommendation\\}
\author{
	Dazhong Rong\IEEEauthorrefmark{2}, 
	Shuai Ye\IEEEauthorrefmark{3}, 
	Ruoyan Zhao\IEEEauthorrefmark{3}, 
	Hon Ning Yuen\IEEEauthorrefmark{2},
	Jianhai Chen\IEEEauthorrefmark{1}\IEEEauthorrefmark{2} \thanks{\IEEEauthorrefmark{1} Corresponding author},
	Qinming He\IEEEauthorrefmark{2}\\
	\IEEEauthorrefmark{2}College of Computer Science and Technology, Zhejiang University, Hangzhou, China\\
	\IEEEauthorrefmark{3}Polytechnic Institute, Zhejiang University, Hangzhou, China\\
	\{rdz98, zjuyeh, ryanzh, yuenhn, chenjh919, hqm\}@zju.edu.cn\\
}
\maketitle

\begin{abstract}
Federated Recommendation (FR) has received considerable popularity and attention in the past few years.
In FR, for each user, its feature vector and interaction data are kept locally on its own client thus are private to others.
Without the access to above information, most existing poisoning attacks against recommender systems or federated learning lose validity.
Benifiting from this characteristic, FR is commonly considered fairly secured.
However, we argue that there is still possible and necessary security improvement could be made in FR.
To prove our opinion, in this paper we present FedRecAttack, a model poisoning attack to FR aiming to raise the exposure ratio of target items.
In most recommendation scenarios, apart from private user-item interactions (\textit{e.g.,} clicks, watches and purchases), some interactions are public (\textit{e.g.,} likes, follows and comments).
Motivated by this point, in FedRecAttack we make use of the public interactions to approximate users' feature vectors, thereby attacker can generate poisoned gradients accordingly and control malicious users to upload the poisoned gradients in a well-designed way.
To evaluate the effectiveness and side effects of FedRecAttack, we conduct extensive experiments on three real-world datasets of different sizes from two completely different scenarios.
Experimental results demonstrate that our proposed FedRecAttack achieves the state-of-the-art effectiveness while its side effects are negligible.
Moreover, even with small proportion ($3\%$) of malicious users and small proportion ($1\%$) of public interactions, FedRecAttack remains highly effective, which reveals that FR is more vulnerable to attack than people commonly considered.
\end{abstract}

\begin{IEEEkeywords}
Recommender System, Federated Recommendation, Federated Learning, Poisoning Attack
\end{IEEEkeywords}
\section{Introduction}
Research in the field of recommender systems mainly focused on boosting recommendation accuracy in the past few years.
Specifically, the studies can be roughly divided into three categories.
The first stream is to construct more sophisticated structures of recommender models (\textit{e.g.,} NCF~\cite{he2017neural}, NAIS~\cite{nais}, NGCF~\cite{wang2019neural} and LightGCN~\cite{he2020lightgcn}).
The second stream is to integrate more auxiliary information (\textit{e.g.,} VBPR~\cite{he2016vbpr}, CKE~\cite{zhang2016collaborative}, MMGCN~\cite{wei2019mmgcn} and KGIN~\cite{intents}).
The third stream is to adapt recommender models to scenarios of more specific areas (\textit{e.g.,} e-commerce~\cite{wang2018billion}, news~\cite{wang2018dkn,zheng2018drn} and micro-videos~\cite{wei2019personalized}).
Note that combining knowledge graph to improve model expressivity is the current research trend~\cite{liu2021combining}.

Up to recent years, the performance of recommender systems was already quite advanced.
As recommender systems were widely used in diverse scenarios, researchers' attention shifted to the security of recommender systems.
Security issues are relevant to everyone, as recommender systems play an important role in people's ways of thinking, judging and acquitting information.
An attacker could make his specific target items to be popular if he gains control of a recommender system.
In severe cases, attacker can even influence public opinion on social media and change the public's views or attitudes on specific heat events, resulting in changes in people's actual behaviors. 
Cambridge Analytica is a data analysis company which manipulates political public opinion.
They capture user-item interactions on different social media and use the data to generate user profiles, then accurately recommend particular advertisements to specific users.
As a result, the company has a direct and decisive impact on the results of American Presidential Election~\cite{gonzalez2017hacking}.

References~\cite{li2016data, huang2021data, fang2018poisoning} present data poisoning attacks against matrix factorization based, deep learning based and graph based recommender systems respectively.
All of the above attacks highly rely on attacker's prior knowledge of user-item interactions to approximate the parameters of recommender model, thus attacker can control malicious users to generate fake interactions accordingly, resulting in the exposure ratio of target items being raised.

Meanwhile, people were also paying increasing attention to privacy protection.
In traditional centralized recommender systems, servers store all historical user-item interactions and private profiles of millions of users, which contain vast amount of sensitive information.
Therefore, once there is leakage in data on servers, there will be extremely serious negative impact on millions of users, enterprises and the society. 
For example, \textit{Wired}, \textit{The New York Times} and \textit{The Observer} reported that Facebook had been repeatedly suffering from server data breaches in recent years.
In the most severe case, data of more than 500 million users was exposed, causing pensions of 5 billion dollars to Facebook~\cite{chauhan20212021}.
In 2016, European Union promulgated the General Data Protection Regulation (GDPR), which prohibited commercial companies from collecting, processing or exchanging user data without the corresponding user’s permission~\cite{regulation2018general}.
Afterwards, the United States, China, Egypt and Brazil also formulated and introduced similar regulations to protect users' privacy~\cite{kalyanpur2019mnc}.

\begin{figure*}
	\setlength{\abovecaptionskip}{-20pt}
	\centering
	\includegraphics[width=0.83\linewidth]{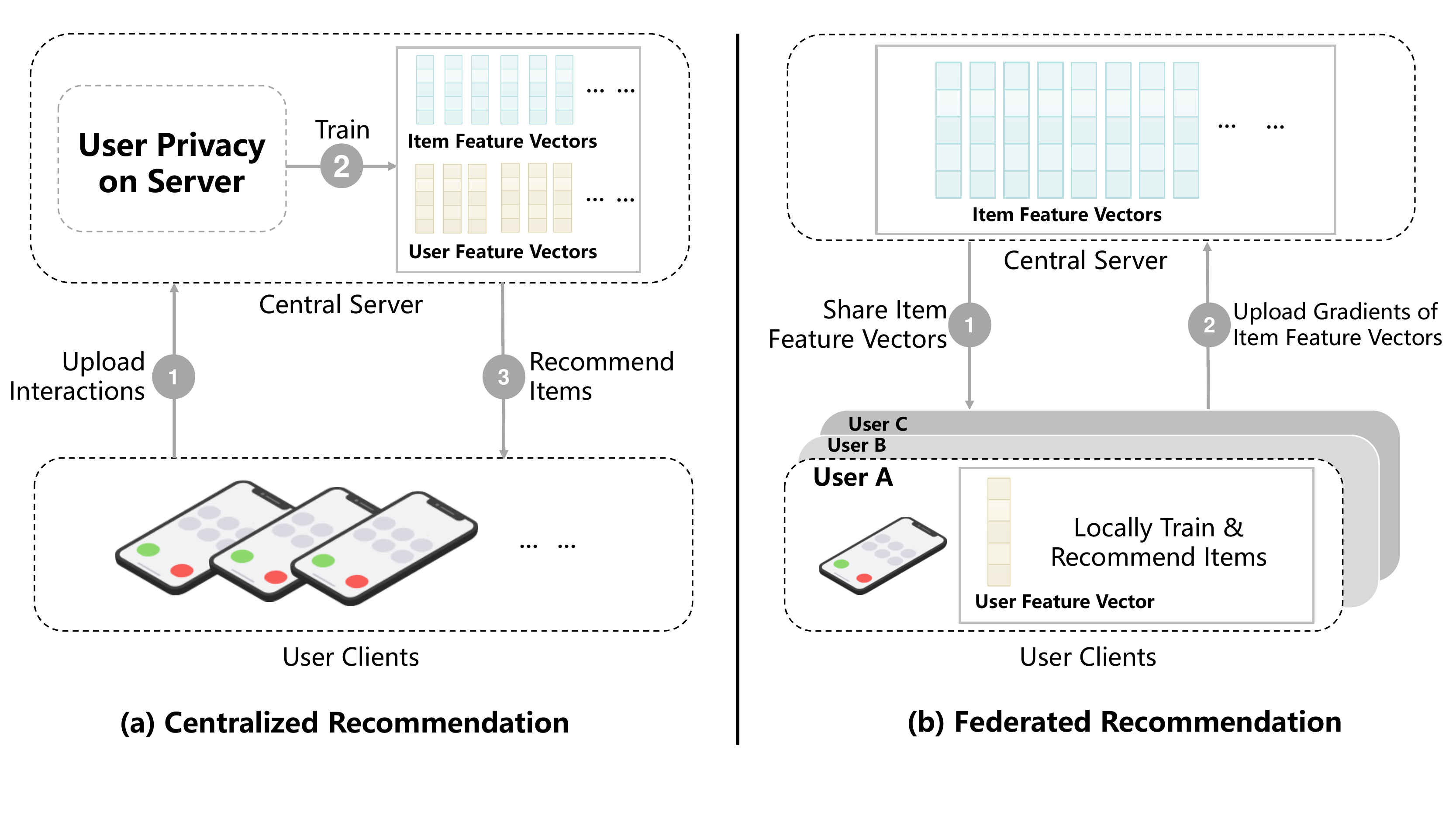}
	\caption{Centralized Recommendation v.s. Federated Recommendation}
	\label{fig:shuangta}
\end{figure*}

Federated recommendation (FR) is considered as a solution to recommend items under circumstances of protecting users' privacy.
In FR, recommender systems are trained in federated scenarios.
As~\cite{ammad2019federated} proposed the first framework of FR, other researches followed their work and raised extended versions~\cite{lin2020fedrec, liang2021fedrec++, wu2021fedgnn, muhammad2020fedfast, wu2021hierarchical, minto2021stronger} in a short time.

The strengths of FR include not only protecting users' privacy, but also being immune to existing attacks against recommender systems.
As shown in Fig.~\ref{fig:shuangta}, different from centralized recommendation, in FR for each user its interaction data and feature vector are kept locally on its own client thus are private to others (including central server).
This means that it is impractical for attacker to access users' interaction data.
Because of the dependency on users' interaction data, existing data poisoning attacks lose validity in FR.
Similarly, existing model poisoning attacks against federated learning~\cite{bhagoji2019analyzing,bagdasaryan2020backdoor,fang2020local} are also inapplicable in FR.
The reason is, all of these model poisoning attacks require full access to model parameters.
However, each user's feature vector is also private in FR.
Based on the above evidences, people consider FR fairly secured, therefore it could be stated that attacking federated recommender systems is a highly challenging task.

Nevertheless, we believe that there is still possible and necessary improvement to the security of FR.
On numerous social platforms (\textit{e.g.,} twitter, facebook and instagram), apart from private interactions (\textit{e.g.,} clicks, watches and purchases), some interactions are public (\textit{e.g.,} likes, follows and comments).
Although the proportion of public interactions is usually small, it could still make significant contributions to attacks with the help of shared model parameters in FR.
Inspired by the above hypothesis, we propose FedRecAttack.
In FedRecAttack, attacker can approximate users' feature matrix with public interactions and the shared parameters.
Subsequently, attacker can generate poisoned gradients accordingly, and control malicious users to upload them in a well-designed way.
The \textbf{key challenges} of our work are summarized below.

\textbf{Conducting attacks with only small proportion of public interactions.}\quad
Existing data poisoning attacks against recommender systems highly rely on users' interaction data, which is inaccessible for attacker in FR.
Even with the help of public interactions, these attacks remain ineffective, due to the low proportion.
In FedRecAttack, we need to devise an ingenious way to make adequate use of public interactions.

\textbf{Conducting attacks without access to users' feature matrix.}
In FR, model parameters can be separated into two parts, one is shared while the other is private.
The shared model parameters are maintained on central server.
In contrast, the private model parameters are kept locally on each user client.
Existing model poisoning attacks agaist federated learning require full access to both parts, which is obviously impractical.
In FedRecAttack, we need to validly approximate the private model parameters (\textit{i.e.,} users' feature matrix) with the help of public interactions and the shared model parameters.

\textbf{Conducting attacks by stealth.}
To the best of our knowledge, PipAttack~\cite{zhang2021pipattack} is the first proposed attack method against FR.
However, there are two drawbacks of PipAttack.
The first is it requires attacker's prior knowledge of side information about items' popularity, which is not always accessible in FR. 
The second is it leads to significant degradation of recommendation accuracy, indicating the attack is highly detectable.
To ensure both the stealthiness and the effectiveness of FedRecAttack, we need to achieve attacker's goal without leading significant side effects.

In this paper, we attempt to explore the possibility of attacking federated recommender systems. 
We devise an ingenious way to approximate users' feature matrix and use it to compute poisoned gradients.
Additionally, we design a sophisticated method to upload poisoned gradients under certain limitations.
Our attack can poison the recommender model effectively, while the side effects of our attack are negligible, which demonstrates the vulnerability of FR. 

The \textbf{key contributions} of our work can be summarized as following:
\begin{itemize}
	\item
	We propose FedRecAttack, the first open source model poisoning attack method against FR.\footnote{Our implementation is available at https://github.com/rdz98/FedRecAttack}
	As evaluated on three real-world datasets of different sizes from two completely different scenarios, our FedRecAttack reaches the highest effectiveness among all baseline attacks.
	
	\item
	We compare the effectiveness of FedRecAttack with that of state-of-the-art data poisoning and model poisoning attacks.
	The effectiveness of FedRecAttack sets the new state-of-the-art when the proportion of malicious users is small.
	
	\item
	We explore the side effects of FedRecAttack.
	The experimental results demonstrate that the side effects of FedRecAttack are the slightest in all model poisoning attacks, which means FedRecAttack is difficult to be detected.
\end{itemize}

\section{Related Work}
\subsection{Recommender Systems}
Collaborative filtering recommender systems use user-item interactions (\emph{e.g.,} clicks or favors) to model the latent features of users and items. 
Training and optimizing the recommender model with historical user-item interactions, empowers the recommender system to predict rating score between uninteracted user-item pair, describing how much the user likes the item.
Previous researches on Recommender Systems (RS) can be categorized into three groups:
i) Matrix factorization based RS~\cite{mnih2007probabilistic,kabbur2013fism,kumar2014social,he2016fast}, which simply take the dot product of user's and item's feature vectors as the predicted rating score between the user and the item.
ii) Deep learning based RS~\cite{he2017neural,bai2017neural,he2018outer,sedhain2015autorec,wu2016collaborative,xue2017deep}, which use deep learning models to make the prediction of rating scores with high level of non-linearities.
iii) Graph based RS~\cite{wang2019neural,he2020lightgcn}, which consider user-item interactions as the edges in a bipartite graph and try to exploit the high-order connectivity by performing embedding propagation.

All these researches designed elaborate model structures and boosted the recommendation performance.

\subsection{Attacks against Recommender Systems}
Fueled by the technical maturity, RS have become more widely used in diverse scenarios continuously, along with the number of researchers focusing on security issues of RS growing.
Existing studies show that current RS are vulnerable under certain circumstances as researchers proposed many effective attack methods against RS.
One line of the attacks aim to cause loss of the recommendation accuracy and undermine the validity of target model.
Another line of the attacks aim to make specific items to be recommended to as many users as possible.

In this paper, we focus on the latter attacks as they are more stealthy and more harmful.
Data poisoning attack is the main form of the attacks, which proceeds by injecting fake users and controlling them to interact with some carefully selected items, leading to training data being poisoned.
The recommender model trained on the poisoned data will predict abnormally high rating scores of the target items to many users.
References~\cite{li2016data,fang2018poisoning,fang2020influence,huang2021data} proposed data poisoning attacks against the above three classes of RS.
Nevertheless, all of these attacks have limited effects and highly rely on attacker's prior knowledge (\emph{i.e.,} historical interactions).
More specifically, in~\cite{li2016data,fang2018poisoning} attacker was assumed to have access to full interactions, and in~\cite{fang2020influence, huang2021data} attacker was assumed to have access to partial interactions (more than $20\%$).
These assumptions are often impractical in real-world scenarios.

\subsection{Federated Recommendation}
Federated Recommendation (FR) is considered to be immune to such attacks.
Fig.~\ref{fig:shuangta} shows the differences between the architectures of traditional centralized recommendation and FR.
In FR, for each user, its interaction data is kept locally on its own client.
As mentioned above, attacker's prior knowledge about historical interactions is critical to the attacks, which means that all these attacks are invalid in FR.

Due to the advantages of protecting users' privacy and immunity from attacks, FR has become a hot research topic.
As~\cite{ammad2019federated} proposed the first framework of FR with implicit feedback, subsequent studies have sprung up.
In~\cite{lin2020fedrec}, the deficiency that explicit feedback is inapplicable in FR was remedied by randomly sampling the items which user did not interact with.
Based on~\cite{lin2020fedrec},~\cite{liang2021fedrec++} proposed a lossless federated recommender system, which used denoising clients to collect the gradients with noise from ordinary clients.
Reference~\cite{wu2021fedgnn} applied graph neural network to FR, and achieved fairly good recommendation accuracy without losing the protection of users' privacy.

\subsection{Attacks against Federated Learning}
Since the advent of Federated Learning (FL), the security issues of FL have been of concern to researchers, and a large number of attacks against FL have been proposed.
Model poisoning attack is the main form of these attacks, which proceeds by controlling malicious users to upload poisoned gradients.
References~\cite{bhagoji2019analyzing,bagdasaryan2020backdoor,fang2020local} presented model poisoning attacks against FL in classification tasks.
FR is a special case of FL.
However existing attacks against FL are invalid in FR because of the following reasons:
i) Most existing attacks are designed in classification tasks, but in FR the learning objective is completely different from that in classification tasks.
ii) Existing attacks rely on attacker's full access to model parameters, but in FR each user's feature vector is kept hidden to others.

\subsection{Attacks against Federated Recommendation}
The attack methods against FR are still under-explored.
Reference~\cite{zhang2021pipattack} proposed PipAttack, the first framework of model poisoning attack against FR.
Its experimental results show that PipAttack is effective in FR.
However, there are two disadvantages of PipAttack:
i) PipAttack causes significant degradation of recommendation accuracy, which increases the probability of being detected.
ii) PipAttack demands a large proportion ($10\%$) of malicious users to achieve its effectiveness, raising the attack costs to a fairly unaffordable extent.

Although both of the above characteristics imply that it is still difficult to cause serious damages to current federated recommender systems, we consider that there is still necessary improvement to the security of FR.
To prove that it is possible to carry out attacks in FR with negligible side effects and small proportion of malicious users, we propose FedRecAttack.

\section{Problem Formulation}
In this section, we first formally define the base recommender and the framework of FR.
Then we introduce the fundamental settings of our attack.

\begin{table}
	\renewcommand\arraystretch{1.3}
	\caption{List of Important Notations}
	\centering
	\begin{tabular}{c|l}
		\hline
		$\boldsymbol{\mathcal{D}}$ & all interactions\\
		$\boldsymbol{\mathcal{D}}'$ & public interactions\\
		\hline
		$$
		$\boldsymbol{\mathcal{V}}_i^+$ & items user $u_i$ has interacted with\\
		$\boldsymbol{\mathcal{V}}_i^-$ & items user $u_i$ has not interacted with\\
		${\boldsymbol{\mathcal{V}}_i^-}'$ & negative items for $u_i$\\
		${\boldsymbol{\mathcal{V}}_i^-}''$ & items user $u_i$ has not publicly interact with\\
		$\boldsymbol{\mathcal{V}}_i^{rec}$ & items in $\boldsymbol{\mathcal{V}}_i^-$ with top-$K$ predicted scores for user $u_i$\\
		${\boldsymbol{\mathcal{V}}_i^{rec}}'$ & items in ${\boldsymbol{\mathcal{V}}_i^-}''$ with top-$K$ predicted scores for user $u_i$\\
		$\boldsymbol{\mathcal{V}}^{tar}$ & target items\\
		\hline
		$\boldsymbol{\mathcal{U}}^t$ & selected benign users in the $t$-th iteration\\
		$\boldsymbol{\mathcal{U}}_m^t$ & selected malicious users in the $t$-th iteration\\
		\hline
		$\nabla\mathbf{V}^{t}_i$ & the gradients of $\mathbf{V}$ uploaded by user $u_i$ in the $t$-th iteration\\
		$\nabla\mathbf{\tilde{V}}^t_i$ & $\nabla\mathbf{V}^{t}_i$ when user $u_i$ is malicious\\
		$\nabla\mathbf{\tilde{V}}^{t}$ & the sum of poisoned gradients of $\mathbf{V}$ in the $t$-th iteration\\
		$\nabla\mathbf{v}^t_{ij}$ & the $j$-th row of $\nabla\mathbf{V}^{t}_i$\\	
		$\nabla\mathbf{\tilde{v}}^t_{ij}$ & the $j$-th row of $\nabla\mathbf{\tilde{V}}^{t}_i$\\	
		$\nabla\mathbf{\tilde{v}}^t_j$ & the $j$-th row of $\nabla\mathbf{\tilde{V}}^{t}$\\
		\hline
		$k$ & dimension of feature vectors\\
		$\mu$ & noise scale\\
		$\eta$ & learning rate\\
		$\xi$ & the proportion of public interactions\\
		$\rho$ & the proportion of malicious users\\
		$\kappa$ & the maximum number of non-zero rows in uploaded gradients\\
		$C$ &  the maximum $\ell_2$-norm of rows in uploaded gradients\\
		\hline
	\end{tabular}
	\label{table-parameter-define}
\end{table}

\subsection{Base Recommender}
We define a target recommender system to contain a set of users $\boldsymbol{\mathcal{U}}=\{u_1, u_2, \dots, u_n\}$ and a set of items $\boldsymbol{\mathcal{V}}=\{v_1, v_2, \dots, v_m\}$, where $n$ and $m$ are the numbers of users and items respectively. 
Training dataset $\boldsymbol{\mathcal{D}}\subseteq \boldsymbol{\mathcal{U}} \times \boldsymbol{\mathcal{V}}$ is composed of implicit feedback tuples which represent user-item interactions (\emph{e.g.,} clicked, watched).
Tuple $(u_i, v_j)\in \boldsymbol{\mathcal{D}}$ indicates the interaction between user $u_i$ and item $v_j$. 
For each user $u_i$, we also define $\boldsymbol{\mathcal{V}}_i^+=\{v\in \boldsymbol{\mathcal{V}}\colon (u_i, v)\in \boldsymbol{\mathcal{D}}\}$ and $\boldsymbol{\mathcal{V}}_i^-=\{v\in \boldsymbol{\mathcal{V}}\colon (u_i, v)\notin \boldsymbol{\mathcal{D}}\}$ as the set of the items user $u_i$ has interacted with and that of those user $u_i$ has not, for convenience.

Let $\mathbf{U}\colon |\boldsymbol{\mathcal{U}}|\times k$ and $\mathbf{V}\colon |\boldsymbol{\mathcal{V}}|\times k$ denote the feature matrices with $k$ columns of users and items respectively.
The feature vectors $\mathbf{u}_i$ and $\mathbf{v}_j$ are the $i$-th row in $\mathbf{U}$ and the $j$-th row in $\mathbf{V}$, describing the latent features of user $u_i$ and item $v_j$ respectively.
The predicted rating score $\hat{x}_{ij}$, which represents how much user $u_i$ likes item $v_j$, is computed as following:
\begin{equation}
	\hat{x}_{ij}=\Upsilon(\mathbf{u}_i, \mathbf{v}_j),
\end{equation}
where $\Upsilon$ is the interaction function.

In matrix factorization based recommender models (\textit{e.g.,} MF~\cite{koren2009matrix}, FISM~\cite{kabbur2013fism}), $\Upsilon$ is fixed.
In deep learning based recommender models (\textit{e.g.,} NCF~\cite{he2017neural}, ONCF~\cite{he2018outer}), $\Upsilon$ is learnable.
Various models are adopted in different scenarios of FR.
It is worth mentioning that our attack is applicable to almost all recommender models as long as they are collaborative filtering.

To be specific, in this paper we adopt Matrix Factorization (MF), one of the most classic and widely used recommender models, as our base recommender.
In MF, $\Upsilon$ is fixed to be dot product (\textit{i.e.,} $\hat{x}_{ij}=\mathbf{u}_i\odot\mathbf{v}_j$, where $\odot$ denotes dot product).

We adopt Bayesian Personalized Ranking (BPR)~\cite{rendle2012bpr} loss to train our base recommender.
The BPR loss for each user is defined as following:
\begin{equation}
	\mathcal{L}_i^{rec}=-\sum_{v_j\in \boldsymbol{\mathcal{V}}_i^+ \land v_k\in \boldsymbol{\mathcal{V}}_i^-}{\ln{\sigma(\hat{x}_{ijk})}}
	,\label{single-user-bpr-loss-equation}
\end{equation}
where $\sigma$ is the logistic sigmoid and $\hat{x}_{ijk}=\hat{x}_{ij}-\hat{x}_{ik}$.
The model is optimized by minimizing the BPR loss for all users:
\begin{equation}
	\mathcal{L}^{rec}=\sum_{u_i\in \boldsymbol{\mathcal{U}}}\mathcal{L}_i^{rec}
	.\label{Recommend-Loss-For-All-User}
\end{equation}

In real-world scenarios, various loss functions are applied due to the feedback from users could be implicit or explicit. 
Once again note that our attack is applicable while other popular loss functions (\emph{e.g.,} cross-entropy loss, hinge loss) are adopted.

\subsection{Framework of Federated Recommendation}
We divide the parameters of our base recommender into users' feature matrix $\mathbf{U}$, items' feature matrix $\mathbf{V}$ and other learnable model parameters (denoted by $\mathbf{\Theta}$).
If the interaction function $\Upsilon$ is fixed, $\mathbf{\Theta}$ is an empty set.
If $\Upsilon$ is learnable through a deep neural network, $\mathbf{\Theta}$ is the set of the parameters in the neural network.

In FR there is a central server and a large amount of user clients.
The shared parameters $\mathbf{V}$ and $\mathbf{\Theta}$ are maintained by the central server, while for each user $u_i$ its interaction data $\boldsymbol{\mathcal{V}}_i^+$ and its private parameter $\mathbf{u}_i$ (\textit{i.e.,} the $i$-th row in $\mathbf{U}$) are kept locally on its own client, which is the main difference between federated and centralized recommendation.

In the beginning, the central server initializes the shared parameters $\mathbf{V}$ and $\mathbf{\Theta}$, meanwhile each user client $u_i$ initializes its private parameters $\mathbf{u}_i$.
Moreover, to reduce the complexity of computation, each user client $u_i$ randomly samples a subset of negative items (denoted by ${\boldsymbol{\mathcal{V}}_i^-}'$) from ${\boldsymbol{\mathcal{V}}_i^-}$, and uses ${\boldsymbol{\mathcal{V}}_i^-}'$ instead of ${\boldsymbol{\mathcal{V}}_i^-}$.
We can represent $\boldsymbol{\mathcal{V}}_i^+ = \{v_{i1}^+, v_{i2}^+, \dots, v_{ip}^+\}$ and ${\boldsymbol{\mathcal{V}}_i^-}' = \{v_{i1}^-, v_{i2}^-, \dots, v_{ip}^-\}$, where $p=|{\boldsymbol{\mathcal{V}}_i^-}'|=|\boldsymbol{\mathcal{V}}_i^+|$.
The loss function for user $u_i$ can be simplified from Eq.~\eqref{single-user-bpr-loss-equation} to:
\begin{equation}
	\mathcal{L}_i^{rec}=-\sum_{(v_j, v_k)\in \boldsymbol{\mathcal{V}}_i}{\ln{\sigma(\hat{x}_{ijk})}}
	,\label{Recommend-Loss-For-Each-User}
\end{equation}
where $\boldsymbol{\mathcal{V}}_i=\{(v_{i1}^+, v_{i1}^-), (v_{i2}^+, v_{i2}^-), \dots (v_{ip}^+, v_{ip}^-)\}$.

Subsequently, the training stage starts.
In each training iteration, the central server randomly selects a batch of user clients $\boldsymbol{\mathcal{U}'}\subseteq \boldsymbol{\mathcal{U}}$ and sends the shared model parameters $\mathbf{V}$ and $\mathbf{\Theta}$ to them.
For each selected user client $u_i\in \boldsymbol{\mathcal{U}'}$, with its loss $\mathcal{L}_i^{rec}$ computed, it derives the gradients $\nabla\mathbf{V}_i$, $\nabla\mathbf{\Theta}_i$ and $\nabla \mathbf{u}_i$ of parameters $\mathbf{V}$, $\mathbf{\Theta}$ and $\mathbf{u}_i$ respectively.
To strongly protect users' sensitive data, as in~\cite{wei2020federated}, each selected user client $u_i$ adds some noise to 
both $\nabla\mathbf{V}_i$ and $\nabla\mathbf{\Theta}_i$:
\begin{equation}
\begin{split}
	\nabla \mathbf{V}_i\leftarrow\nabla \mathbf{V}_i+\mathcal{N}(\textbf{0}, \mu^2 C^2 \mathbf{I}),\\
	\nabla\mathbf{\Theta}_i\leftarrow\nabla\mathbf{\Theta}_i+\mathcal{N}(\textbf{0}, \mu^2 C^2 \mathbf{I}),
\end{split}\label{Eq-add-noise}
\end{equation}
where $\mathcal{N}$, $\mu$ and $C$ denote normal distribution, the noise scale and the $\ell_2$-norm bound of gradients, respectively.
Afterwards, each selected user client $u_i$ uploads $\nabla\mathbf{V}_i$ and $\nabla\mathbf{\Theta}_i$ to the central server and updates its private model parameter $\mathbf{u}_i$ locally as:
\begin{equation}
	\mathbf{u}_i\leftarrow \mathbf{u}_i - \eta \cdot \nabla \mathbf{u}_i,
\end{equation}
where $\eta$ is the learning rate. 
After collecting the gradients from all user clients in $\boldsymbol{\mathcal{U}'}$, the central server updates $\mathbf{V}$ and $\mathbf{\Theta}$ batchly:
\begin{equation}\begin{split}
	\mathbf{V}\leftarrow\mathbf{V} - \eta \cdot \sum_{u_i\in \boldsymbol{\mathcal{U}'}}{\nabla \mathbf{V}_i},\\
	\mathbf{\Theta}\leftarrow\mathbf{\Theta} - \eta \cdot \sum_{u_i\in \boldsymbol{\mathcal{U}'}}{\nabla \mathbf{\Theta}_i}.
	\label{Eq-Update-V}
\end{split}\end{equation}

Throughout the training stage, all users' privacy is well protected.
As for a user client $u_i$, through its uploaded gradients, the central server can only extract the knowledge of whether an item $v$ is in $(\boldsymbol{\mathcal{V}}_i^+\cup{\boldsymbol{\mathcal{V}}_i^-}')$, but not the knowledge of whether an item $v$ is in $\boldsymbol{\mathcal{V}}_i^+$.

\subsection{Fundamental Settings of Attack}
\label{prior-knowledge-limitation}
\textbf{Attacker's Goal}\quad
For each user $u_i$, we suppose the recommender system recommends $K$ items in $\boldsymbol{\mathcal{V}}_i^-$ with the top-$K$ predicted scores.
Let $\boldsymbol{\mathcal{V}}_i^{rec}$ denote the set of the $K$ items recommended for $u_i$, and let $\boldsymbol{\mathcal{V}}^{tar}$ denote the set of the target items.
The exposure ratio at $K$~\cite{zhang2021pipattack} of the target items is defined as:
\begin{equation}
	\text{ER@K} = \frac{1}{|\boldsymbol{\mathcal{U}}|}\sum_{u_i\in \boldsymbol{\mathcal{U}}}
	{
		\frac{|\boldsymbol{\mathcal{V}}^{tar}\land \boldsymbol{\mathcal{V}}_i^{rec}|}{|\boldsymbol{\mathcal{V}}^{tar}\land \boldsymbol{\mathcal{V}}_i^-|}
	}
	.\label{Eq-ER-K}
\end{equation}
We adopt $\text{ER@K}$ as the metric to evaluate the effectiveness of our attack.
Attacker's goal is to make the target items appear in the top-$K$ recommendation lists of as many users as possible.
In other words, attacker aims to raise $\text{ER@K}$ as high as possible. 

\textbf{Attacker's Prior Knowledge} \quad
In FR, each user' feature vectors and historical interactions are private to others (so as to attacker).
However, as mentioned in the previous section, in many cases a small part (no more than $5\%$) of the whole interactions is publicly available.
Our work attempts to exploit these few public interactions for attacker to estimate the overall distribution of all users' actual feature vectors (\textit{i.e.,} $\mathbf{U}$).
Hence, based on the estimation of $\mathbf{U}$, attacker can compute the poisoned gradients of $\mathbf{V}$ and control the malicious user clients to upload the poisoned gradients to the central server.
Following the common settings of FR, we assume that attacker knows the model structure and some hyper parameters (\emph{e.g.,} learning rate $\eta$) used in the federated system.
Due to the fact that central server shares the model parameter $\mathbf{V}$ and $\mathbf{\Theta}$ at the beginning of each iteration,
once any malicious user client is selected to participate in training, attacker knows the current value of $\mathbf{V}$ and $\mathbf{\Theta}$.
In addition, we assume attacker knows the public part of interactions.
Let $\xi$ denote the propotion of the public interactions.
In our experimental section, we explore the effectiveness of our attack in the case of $\xi=1\%,2\%,3\%,5\%,10\%$ respectively.
Note that all other information (\emph{e.g.,} any benign user's feature vector) is unknown to attacker.

\textbf{Limitations on Attacker}\quad
Considering the circumstances in real scenarios, our attack is conducted under certain restrictions, including:
\begin{itemize}
	\item {
\textbf{The proportion of malicious users, denoted as $\boldsymbol{\rho}$.}
Obviously, the more number of malicious users, the more effective the attack will be.
If there is no restriction on $\rho$, attacker can inject a huge amount of malicious users and simply control them to interact with target items, resulting in rising the popularity of target items easily.
However, too many malicious users will lead to an unaffordable cost of the attack.
To make our work more challenging and practical, we restrict $\rho\le10\%$. 
In our experimental section, we explore our attack effectiveness in the case of $\rho=1\%,2\%,3\%,5\%,10\%$ respectively.
}
	\item {
\textbf{The maximum number of non-zero rows in $\boldsymbol{\nabla\mathbf{V}_i}$, denoted as $\boldsymbol{\kappa}$.}
A non-zero row means a row containing at least one non-zero element.
The more number of non-zero rows in the uploaded poisoned gradients, the more powerful the gradients will be.
If the number is unlimited, the attack is easy to be effective, because each malicious user client $u_i$ could upload $\nabla\mathbf{V}_i$ composed entirely of non-zero rows, which would affect all items' feature vectors.
However, an excessive number of non-zero rows in $\nabla\mathbf{V}_i$ will lead to the attack being detected.
To limit malicious users to behave like benign users, in our experiments we set $\kappa=20,40,60,80,100$ respectively.
}	
	\item {
\textbf{The maximum $\boldsymbol{\ell_2}$-norm of rows in $\boldsymbol{\nabla\mathbf{V}_i}$, denoted as $\boldsymbol{C}$.}
If there is no limit on the sizes of rows in $\nabla\mathbf{V}_i$, malicious users can easily poison items' feature vectors by uploading extremely large gradients, which will also lead to the attack being detected.
Therefore, we limit the $\ell_2$-norm of each row in $\nabla\mathbf{V}_i$ not to be higher than $C$.
}
\end{itemize}

\begin{figure*}[t]
	\setlength{\abovecaptionskip}{-25pt}
	\centering
	\includegraphics[width=0.75\linewidth]{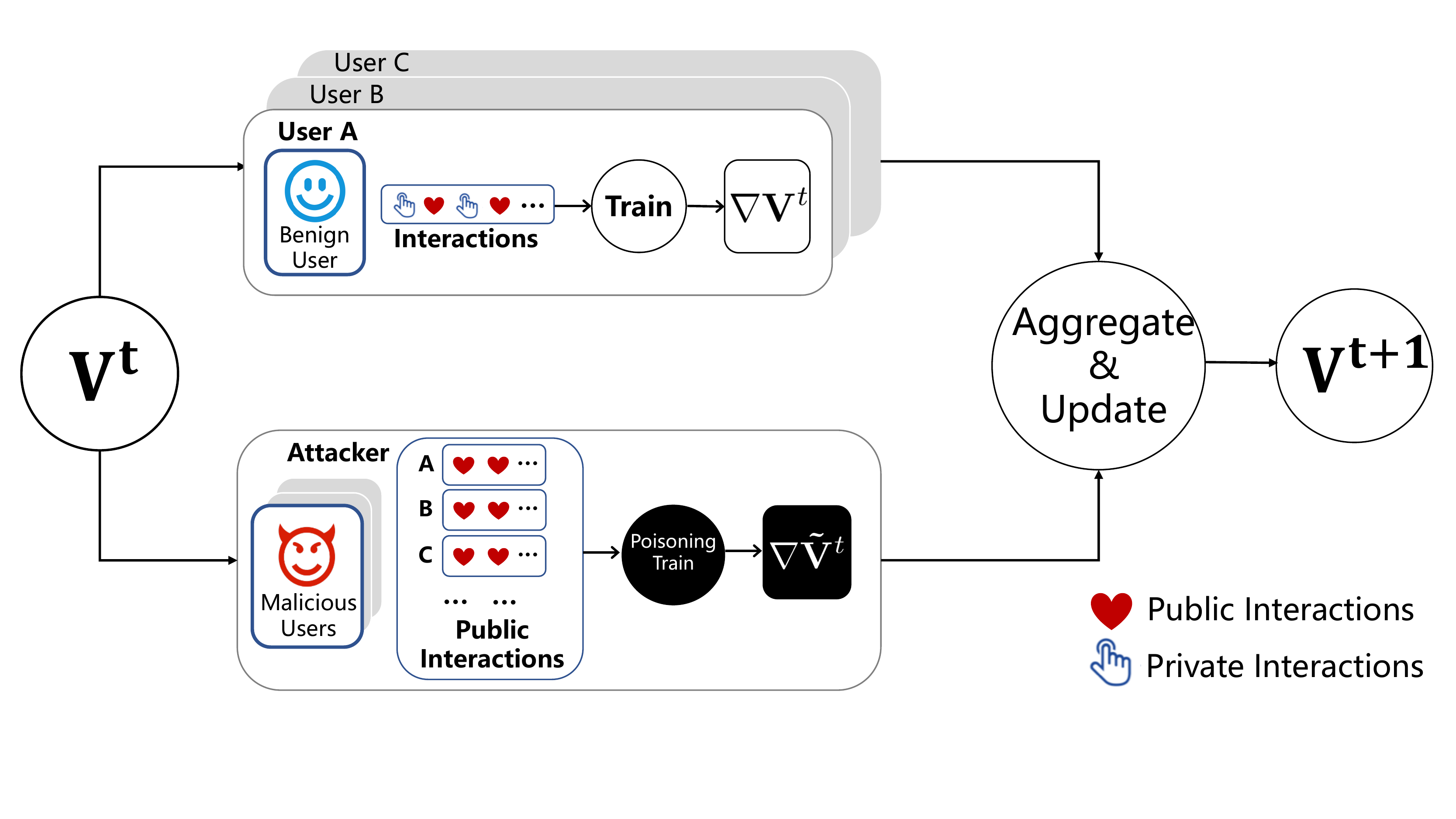}
	\caption{Diagram of FedRecAttack}
	\label{fig:attack}
\end{figure*}
\section{Our Attack: FedRecAttack}

In this section, we present our FedRecAttack, a method of model poisoning attack in FR scenarios.
In FedRecAttack, attacker first computes the poisoning gradients of items' feature matrix $\mathbf{V}$ with the access to the shared model parameters and the prior knowledge of the public interactions.
Then attacker controls malicious clients to upload the poisoning gradients in a well-designed way.

Note that when the recommender is deep learning based, poisoning the learnable interaction function $\Upsilon$ is possibly a simpler and more effective attack method.
However this method is not generic when the base recommender is matrix factorization based, because in this case the interaction function $\Upsilon$ is fixed.
Therefore, to ensure the generality of our attack, in FedRecAttack we consider to poison items' feature matrix $\mathbf{V}$ only.

\subsection{Solving the Optimization Problem}
\textbf{Formulate the Attack as an Optimization Problem}\quad
Let $T$ denote the total number of training iterations.
Let $\boldsymbol{\mathcal{U}}_m^t$ and $\nabla\mathbf{\tilde{V}}^t$ denote the set of selected malicious users and the sum of thier uploaded gradients in the $t$-th iteration, respectively.
Let $\mathbf{U}^t$, $\mathbf{V}^t$ and $\mathbf{\Theta}^t$ denote the parameters $\mathbf{U}$, $\mathbf{V}$ and $\mathbf{\Theta}$ at the end of the $t$-th iteration, respectively. 
Then we can formally define our attack as an optimization problem:
\begin{equation}
	\begin{split}
		&\max_{\{\nabla\mathbf{\tilde{V}}^1, \nabla\mathbf{\tilde{V}}^2, \cdots, \nabla\mathbf{\tilde{V}}^T\}}
		\quad\text{ER@K}(\mathbf{U}^t, \mathbf{V}^t, \mathbf{\Theta}^t).
		\\
		&\mathop{s.t.}_{ t\in \{1,2,\cdots,T\}, u_i\in \boldsymbol{\mathcal{U}}_m^t}\quad
		\left\{\begin{array}{lc}
			||\nabla \mathbf{v}_{ij}^t||_2\le C,\quad & 
			(\forall v_j\in \boldsymbol{\mathcal{V}})\\
			\\
			\sum\limits_{v_j\in \boldsymbol{\mathcal{V}}}{\delta(\nabla \mathbf{v}_{ij}^t)}\le\kappa,\quad &
		\end{array}\right.
	\end{split}
\end{equation}
where $\nabla \mathbf{v}_{ij}^t$ denotes the gradients of item $v_j$ uploaded by user client $u_i$ in the $t$-th iteration, and $\delta$ is a function as:
\begin{equation}
	\delta(\mathbf{v})=
	\begin{cases}
		1, & ||\mathbf(v)||_2> 0\\
		0, & ||\mathbf(v)||_2= 0.
	\end{cases}
\end{equation}

\textbf{Simplify the Optimization Problem}\quad
Let $\boldsymbol{\mathcal{U}}^t$ and $\nabla \mathbf{V}_i^t$ denote the set of selected benign users and the uploaded gradients from user client $u_i$ in the $t$-th iteration respectively.
Following Eq.~\eqref{Eq-Update-V}, we have:
\begin{equation}
	\mathbf{V}^{t+1}=\mathbf{V}^t - \eta \cdot (\nabla\mathbf{\tilde{V}}^{t+1} + \sum_{u_i\in \boldsymbol{\mathcal{U}^t}}{\nabla \mathbf{V}_i^{t+1}}).
\end{equation}
From the above equation, we can conclude that:
\begin{enumerate}
	\item $\mathbf{V}^{t+1}$ depends on both $\nabla\mathbf{\tilde{V}}^{t+1}$ and $\nabla \mathbf{V}_i^{t+1}$.
	\item $\nabla \mathbf{V}_i^{t+1}$ denotes the gradients of $\mathbf{V}^{t}$, so $\nabla \mathbf{V}_i^{t+1}$ also depends on $\nabla\mathbf{\tilde{V}}^{t}$ and $\nabla \mathbf{V}_i^{t}$.
	\item By combining (1) and (2) recursively, $\mathbf{V}^{t+1}$ depends on $\{\nabla\mathbf{\tilde{V}}^{t+1}, \nabla\mathbf{\tilde{V}}^{t}, \cdots, \nabla\mathbf{\tilde{V}}^{1}\}$.
\end{enumerate}
In summary, the poisoned gradients in the $t$-th iteration (\textit{i.e.,} $\nabla\mathbf{\tilde{V}}^t$) have influences on all subsequent iterations.
Because attacker has no access to benign users' data, attacker is unable to compute $\nabla \mathbf{V}_i^{t+1}$.
Therefore, it is impossible for attacker to predict the influences of $\nabla\mathbf{\tilde{V}}^t$ on subsequent learning.
Due to the above reason, finding the global optimal solution of our optimization problem is difficult.
We have to use a greedy strategy to simplify the problem.
We try to find the current optimal solution instead of the global one, without the consideration of gradients uploaded by benign users.
More specifically, in the $t$-th iteration, we need to solve the optimization problem as following:
\begin{equation}
	\begin{split}
		&\max_{\nabla\mathbf{\tilde{V}}^t}\quad
		\text{ER@K}(\mathbf{U}^{t-1}, \mathbf{V}^{t-1}-\eta\nabla\mathbf{\tilde{V}}^t, \mathbf{\Theta}^{t-1}).
		\\
		&\mathop{s.t.}_{u_i\in \boldsymbol{\mathcal{U}}_m^t}\quad  
		\left\{\begin{array}{lc}
			||\nabla \mathbf{v}_{ij}^t||_2\le C,\quad & 
			(\forall v_j\in \boldsymbol{\mathcal{V}})\\
			\\
			\sum\limits_{v_j\in \boldsymbol{\mathcal{V}}}{\delta(\nabla \mathbf{v}_{ij})}\le\kappa.\quad &
		\end{array}\right.
	\end{split}
	\label{Formulation-of-Optimization-Problem}
\end{equation}

\textbf{Approximate the Exposure Ratio}\quad
The exposure ratio (\textit{i.e.,} $\text{ER@K}$) is a discontinuous and non-differentiable function of predicted scores, thus optimizing it by gradient descent is impossible.
It is still difficult to solve the simplified optimization problem.
To address the challenge, we design a continuous loss function to indicate $\text{ER@K}$ of target items.
For each benign user $u_i$, we define its loss function as:
\begin{equation}
	\mathcal{L}_i^{atk}=\sum_{v_t\in\boldsymbol{\mathcal{V}}^{tar}\land (u_i,v_t)\notin\boldsymbol{\mathcal{D}}}
	{g(\min_{v_j\in\boldsymbol{\mathcal{V}}_i^{rec}\land v_j\notin\boldsymbol{\mathcal{V}}^{tar}}
		\{\hat{x}_{ij}\}-\hat{x}_{it})},
	\label{eq-attack-loss-for-single-user}
\end{equation}
where the function $g$ is defined as:
\begin{equation}
	g(x)=\begin{cases}
		x, & x \ge 0\\
		e^x-1, & x < 0.
	\end{cases}
\end{equation}
Let $\boldsymbol{\mathcal{D}}'\subseteq \boldsymbol{\mathcal{D}}$ ($|\boldsymbol{\mathcal{D}}'|\le\xi |\boldsymbol{\mathcal{D}}|$) denote the set of public interactions which is accessible to attacker.
Then we can define ${\boldsymbol{\mathcal{V}}_i^-}''=\{v\in \boldsymbol{\mathcal{V}}\colon (u_i, v)\notin \boldsymbol{\mathcal{D}}'\}$, and use $\boldsymbol{\mathcal{V}}_i^{rec'}$ to denote the set of items with top-$K$ predicted scores for user $u_i$ in ${\boldsymbol{\mathcal{V}}_i^-}''$.
Due to the limitations on attacker's prior knowledge, $\boldsymbol{\mathcal{D}}$ and $\boldsymbol{\mathcal{V}}_i^{rec}$ are hidden from attacker, thus we have to approximate them with $\boldsymbol{\mathcal{D}}'$ and $\boldsymbol{\mathcal{V}}_i^{rec'}$ respectively.
The loss function for each user is now transformed to be:
\begin{equation}
	\mathcal{L}_i^{atk}=\sum_{v_t\in\boldsymbol{\mathcal{V}}^{tar}\land (u_i,v_t)\notin\boldsymbol{\mathcal{D}}'}
	{g(\min_{v_j\in\boldsymbol{\mathcal{V}}_i^{rec'}\land v_j\notin\boldsymbol{\mathcal{V}}^{tar}}
		\{\hat{x}_{ij}\}-\hat{x}_{it})}
	,\label{Attack-Loss-For-Each-User}
\end{equation}
The loss function for all benign users is defined as:
\begin{equation}
	\mathcal{L}^{atk}=\sum_{u_i\in \boldsymbol{\mathcal{U}}}
	\mathcal{L}_i^{atk}
	.\label{Attack-Loss-For-All-User}
\end{equation}
Now we can raise $\text{ER@K}$ of target items by minimizing $\mathcal{L}^{atk}$.

\textbf{Approximate Users' Feature Matrix}\quad
Because $\mathcal{L}^{atk}$ depends on the private parameter $\mathbf{U}$, attacker is not able to compute $\mathcal{L}^{atk}$.
To address this challenge, we propose an effective method to approximate $\mathbf{U}$ with the shared model parameters (\textit{i.e.,} $\mathbf{V}$ and $\mathbf{\Theta}$) and the public interactions (\textit{i.e.,} $\boldsymbol{\mathcal{D}}'$), which are accessible to attacker.
Remember that the goal of training a recommender model is to optimize the model parameters $\{\mathbf{U},\mathbf{V},\mathbf{\Theta}\}$ to $\{\mathbf{U}^*,\mathbf{V}^*,\mathbf{\Theta}^*\}$, where:
\begin{equation}
	\{\mathbf{U}^*,\mathbf{V}^*,\mathbf{\Theta}^*\}=\mathop{\arg\min}_{\{\mathbf{U},\mathbf{V},\mathbf{\Theta}\}}
	\{\mathcal{L}^{rec}(\mathbf{U},\mathbf{V},\mathbf{\Theta};\boldsymbol{\mathcal{D}})\}.
\end{equation}
We can further infer that:
\begin{equation}
	\mathbf{U}^*=\mathop{\arg\min}_{\mathbf{U}}
	\{\mathcal{L}^{rec}(\mathbf{U},\mathbf{V}^*,\mathbf{\Theta}^*;\boldsymbol{\mathcal{D}})\}.
\end{equation}
Once again, considering the limitations on attacker's prior knowledge, we replace $\boldsymbol{\mathcal{D}}$ with $\boldsymbol{\mathcal{D}}'$, and approximate $\mathbf{U}^t$ as:
\begin{equation}
	\mathbf{U}^t\approx\mathop{\arg\min}_{\mathbf{U}}
	\{\mathcal{L}^{rec}(\mathbf{U},\mathbf{V}^t,\mathbf{\Theta}^t;\boldsymbol{\mathcal{D}}')\},
	\label{Approximate-Ut}
\end{equation}
Now we can approximate $\mathbf{U}^t$ by using stochastic gradient descent to minimize $\mathcal{L}^{rec}(\mathbf{U},\mathbf{V}^t,\mathbf{\Theta}^t;\boldsymbol{\mathcal{D}}')$.

\textbf{Compute Poisoned Gradients}\quad
Finally, we clearly formulate our attack. 
In the $t$-th training iteration, we need to calculate the best poisoning gradients $\nabla\mathbf{\tilde{V}}^t$ to minimize $\mathcal{L}^{atk}(\mathbf{U}^{t-1},\mathbf{V}^{t-1}-\eta\nabla\mathbf{\tilde{V}}^t,\mathbf{\Theta}^{t-1};\boldsymbol{\mathcal{D}}')$.
We compute $\nabla\mathbf{\tilde{V}}^t$ as following:
\begin{equation}
	\nabla\mathbf{\tilde{V}}^t=\zeta\cdot
	\frac{\partial}{\partial\mathbf{V}}\mathcal{L}^{atk}
	(\mathbf{U}^{t-1},\mathbf{V}^{t-1},\mathbf{\Theta}^{t-1};\boldsymbol{\mathcal{D}}')
	,\label{eq-poisoning-gradient}
\end{equation}
where $\zeta$ is the step size.

\begin{algorithm}[t]
	\caption{Steps of FedRecAttack in the $t$-th training iteration.}\label{algorithm}
	\SetAlgoLined
	\SetKwInOut{Input}{Input}\SetKwInOut{Output}{Output}\SetKwComment{Comment}{/* }{ */}
	\small \Comment{Assuming $p$ malicious users $u_{m_1}$, $u_{m_2}$, $\cdots$, $u_{m_p}$ are selected to participate in the $t$-th training interaction}
	\normalsize
	\Input{the set of public interactions $\boldsymbol{\mathcal{D}}'$\\shared model parameters $\mathbf{V}^t$ and $\mathbf{\Theta}^t$}
	\Output{poisoned gradients $\nabla\mathbf{\tilde{V}}^t_{m_1}$, $\nabla\mathbf{\tilde{V}}^t_{m_2}$, $\dots$, $\nabla\mathbf{\tilde{V}}^t_{m_p}$ for corresponding malicious users to upload in the $t$-th iteration}
	\BlankLine
	
	Approximate $\mathbf{U}^t$ by Eq.~\eqref{Approximate-Ut}\;
	Generate poisoned gradients $\nabla\mathbf{\tilde{V}}^t$ by Eq.~\eqref{eq-poisoning-gradient}\;
	Initialize empty set $\mathbf{G}$ of poisoned gradients.\;
	\ForEach{$u_i$ in $\{u_{m_1}, u_{m_2}, \dots, u_{m_p}\}$}{
		\If{the first time for $u_i$ participating in training}{
			Initialize $\boldsymbol{\mathcal{V}}_i$ by Eq.~\eqref{eq-compute-non-zero-items}
		}
		Compute $\nabla\mathbf{\tilde{V}}^t_i$ by Eq.~\ref{eq-compute-poisoning-gradient}\;
		Control malicious user $u_i$ to upload $\nabla\mathbf{\tilde{V}}^t_i$\;
		Update $\nabla\mathbf{\tilde{V}}^t$ by Eq.~\ref{eq-update-poisoning-gradient}\;
		$\mathbf{G}\leftarrow\mathbf{G}\cup \nabla\mathbf{\tilde{V}}^t_i$
	}
	return $\mathbf{G}$
\end{algorithm}
\subsection{Upload the Poisoned Gradients}
Due to the restrictions on Eq.~\eqref{Formulation-of-Optimization-Problem}, attacker can not upload the poisoned gradients $\nabla\mathbf{\tilde{V}}^t$ directly.
Hence, we design an elaborate method to upload the poisoned gradients.
More specifically, in the $t$-th training iteration, each selected malicious user $u_i\in\boldsymbol{\mathcal{U}}_m^t$ follows the three steps below.

\textbf{Compute $\boldsymbol{\mathcal{V}}_i$} \quad
If it is the first time user $u_i$ to participate in training, we compute $\boldsymbol{\mathcal{V}}_i$ as following:
\begin{equation}
	\boldsymbol{\mathcal{V}}_i=\boldsymbol{\mathcal{V}}^{tar}\cup R(\nabla\mathbf{\tilde{V}}^t,\kappa-|\boldsymbol{\mathcal{V}}^{tar}|)
	,\label{eq-compute-non-zero-items}
\end{equation}
where $R(\nabla\mathbf{\tilde{V}}^t,\kappa-|\boldsymbol{\mathcal{V}}^{tar}|)$ is a function to stochasticly select $(\kappa-|\boldsymbol{\mathcal{V}}^{tar}|)$ items without replacement.
The probability of item $v_j$ being selected is:
\begin{equation}
	p(v_j)=\begin{cases}
		0, & v_j\in\boldsymbol{\mathcal{V}}^{tar}\\
		\\
		\frac{\displaystyle||\nabla\mathbf{\tilde{v}}^t_j||_2}
		{\displaystyle\sum_{v_k\in\boldsymbol{\mathcal{V}}\setminus\boldsymbol{\mathcal{V}}^{tar}}
			{||\nabla\mathbf{\tilde{v}}^t_k||_2}}, 
		& v_j\notin\boldsymbol{\mathcal{V}}^{tar},
	\end{cases}
\end{equation}
where $\nabla\mathbf{\tilde{v}}^t_j$ indicates the $j$-th row in $\nabla\mathbf{\tilde{V}}^t$.
The larger $\nabla\mathbf{\tilde{v}}^t_j$, the higher the probability of item $v_j$ being selected.
Besides,we keep $\boldsymbol{\mathcal{V}}_i$ unchanged if it is not the first time user $u_i$ to participate in training.

\textbf{Compute $\nabla\mathbf{\tilde{V}}^t_i$}\quad
Let $\nabla\mathbf{\tilde{V}}^t_i$ denote the uploaded poisoned gradients from malicious user $u_i$, and $\nabla\mathbf{\tilde{v}}^t_{ij}$ denote the $j$-th row in $\nabla\mathbf{\tilde{V}}^t_i$.
We compute $\nabla\mathbf{\tilde{v}}^t_{ij}$ as following:
\begin{small}
	\begin{equation}
		\nabla\mathbf{\tilde{v}}^t_{ij}=\begin{cases}
			\mathbf{0}, & v_j\notin\boldsymbol{\mathcal{V}}_i\\
			\\
			\nabla\mathbf{\tilde{v}}^t_{j}, 
			& v_j\in\boldsymbol{\mathcal{V}}_i \land ||\nabla\mathbf{\tilde{v}}^t_{j}||_2\le C\\
			\\
			C\cdot \displaystyle\frac{\nabla\mathbf{\tilde{v}}^t_{j}}{||\nabla\mathbf{\tilde{v}}^t_{j}||_2}, 
			& v_j\in\boldsymbol{\mathcal{V}}_i \land ||\nabla\mathbf{\tilde{v}}^t_{j}||_2> C.
		\end{cases}
		\label{eq-compute-poisoning-gradient}
	\end{equation}
\end{small}

\textbf{Update $\nabla\mathbf{\tilde{V}}^t$}\quad
After the computation of $\nabla\mathbf{\tilde{V}}^t_i$, we update $\nabla\mathbf{\tilde{V}}^t$ as following:
\begin{equation}
	\nabla\mathbf{\tilde{V}}^t\leftarrow\nabla\mathbf{\tilde{V}}^t-\nabla\mathbf{\tilde{V}}^t_i
	.\label{eq-update-poisoning-gradient}
\end{equation}

The complete procedure of FedRecAttack is illustrated in Algorithm~\ref{algorithm}.

\section{Experiments}
In this section, firstly, we introduce our experimental settings.
Secondly, we demonstrate the impact of the limitations on attacker in FedRecAttack.
Thirdly, we compare the effectiveness of FedRecAttack with that of other attack methods.
Forthly, we analyze the side effects of FedRecAttack on recommendation accuracy.
And lastly, we conduct an ablation test to demonstrate the necessity of attacker's prior knowledge (\textit{i.e.,} the publicly accessible interactions).


\subsection{Experimental Settings}
\vspace{2px}
\noindent \textbf{Datasets}\quad
We use three real-world datasets in two completely different scenarios (movie recommendation and game recommendation) for our experiments. 
The three datasets are: \textbf{MovieLens-100K}, \textbf{MovieLens-1M}~\cite{harper2015movielens} and \textbf{Steam-200K}~\cite{cheuque2019recommender}.
Their sizes are shown in Table~\ref{table-sizes-of-datasets}.
MovieLens is a dataset of users' ratings for movies.
As a version of small size, MovieLens-100K involves $943$ users, $1,682$ movie items and $100,000$ user-item interactions. MovieLens-1M is an extended version involving $6,040$ users, $3,706$ items and $1,000,209$ interactions.
Steam-200K is a dataset of users' behaviors (own and play) on Steam, the largest and the most famous platform of games.
It involves $3,753$ users, $5,134$ game items and $114,713$ interactions.
In data preprocessing, we transform all kinds of interactions into implicit feedback, and drop the duplicate interactions.
We use the leave-one-out method to divide the training set and test set.
For each user $u_i\in \boldsymbol{\mathcal{U}}$, we randomly select $\xi$ of items in $\boldsymbol{\mathcal{V}}_i^+$, and expose the interactions between user $u_i$ and these selected items to attacker.
\begin{table}[h]
	\centering
	\caption{Sizes of Datasets}
	\resizebox{0.49\textwidth}{!}{
		\begin{tabular}{|c|c|c|c|c|c|}
			\hline
			\textbf{Dataset} & \textbf{\#users} & \textbf{\#items} & \textbf{\#interactions} & \textbf{Avg.} & \textbf{Sparsity}\\
			\hline
			MovieLens-100K & 943   & 1,682 & 100,000 & 106 & 93.70\% \\
			\hline
			MovieLens-1M & 6,040 & 3,706 & 1,000,209 & 166 & 95.53\% \\
			\hline
			Steam-200K & 3,753 & 5,134 & 114,713 & 31 & 99.40\% \\
			\hline
		\end{tabular}%
	}
\label{table-sizes-of-datasets}%
\end{table}%

\vspace{2px}            
\noindent \textbf{Baseline Attacks}\quad 
In our experiments, we compare FedRecAttack with several baseline attack methods, including: 
\begin{itemize}
	\item \textbf{Random Attack}~\cite{gunes2014shilling}: For each malicious user client, attacker randomly selects $(\lfloor\frac{\kappa}{2}\rfloor - |\boldsymbol{\mathcal{V}}^{tar}|)$ items in addition to $\boldsymbol{\mathcal{V}}^{tar}$, and generates fake interactions between the malicious user and the items.
	
	\item \textbf{Bandwagon Attack}~\cite{kapoor2017review}: Similar to random attack but different, attacker randomly selects items based on items' popularity.
	More specifically, we define popular items as the set of the top $10\%$ of items which have the most interactions.
	For each malicious client, in addition to $\boldsymbol{\mathcal{V}}^{tar}$, attacker randomly selects $(\lfloor\frac{\kappa}{2}\rfloor - |\boldsymbol{\mathcal{V}}^{tar}|)\times 10\%$ items from popular items, and $(\lfloor\frac{\kappa}{2}\rfloor - |\boldsymbol{\mathcal{V}}^{tar}|)\times 90\%$ items among the left unselected items.
	
	\item \textbf{Popular Attack} \cite{gunes2014shilling}: In addition to $\boldsymbol{\mathcal{V}}^{tar}$, attacker selects the top $(\lfloor\frac{\kappa}{2}\rfloor - |\boldsymbol{\mathcal{V}}^{tar}|)$ items which have the most interactions. And attacker generates fake interactions between all malicious users and the items.
\end{itemize}
Besides, we use \textbf{None} to indicate the situations when no attacks are conducted.
All the baseline attacks increase $\text{ER@K}$ of target items by raising their popularity and making their feature vectors similar to those of the popular items.

\vspace{2px}    
\noindent \textbf{Evaluation Metrics}\quad
To evaluate attack effectiveness, we adopt three generally-used metrics: $\text{ER@5}$, $\text{ER@10}$ and $\text{NDCG@10}$.
The computation of $\text{ER@5}$ and $\text{ER@10}$ follows Eq.~\eqref{Eq-ER-K}.
For a target item, its rank in user's recommendation list will affect the probability of the item being interacted with by the user.
Hence, we are interested not only in $\text{ER@K}$ of target items, but also in the ranks of target items in users' recommendation lists.
To reflect the ranks, as in~\cite{krichene2020sampled}, we also adopt $\text{NDCG@10}$ as one of our metrics.

\vspace{2px}       
\noindent \textbf{Parameter Setting}\quad 
Unless otherwise specified, we take the default values of the hyper parameters as: $k=32,\eta=0.01,\xi=1\%,\rho=5\%,\kappa=60,C=1,\zeta=1$.
For ensuring recommender model convergence, we set the total number of training epochs to 200.
The attack effectiveness is evaluated at the end of the last training epoch.
We conduct our experiments on a Ubuntu Server with 8 NVIDIA Tesla T4 GPUs, 64-bit 18-core Intel(R) Xeon(R) CPU Gold 6240 @ 2.60GHz and 256 GBs of RAM.

\subsection{Impact of Limitations on Attacker}
To explore the impact of limitations on attacker, we set up three auxiliary comparative experiments on MovieLens-100K with different values of $\xi$, $\rho$ and $\kappa$ respectively.

\textbf{The proportion of public interactions ($\boldsymbol{\xi}$)}\quad
Table~\ref{table-impact-of-prior-knowledge} shows the effectiveness of FedRecAttack with different values of $\xi$ in $\{1\%,2\%,3\%,5\%,10\%\}$.
As can be inferred from the table, increasing the proportion of public interactions can improve attack effectiveness, but the benifits diminish.
When $\xi$ is only $1\%$, our attack is quite effective.
Even if $\xi$ is increased to $10\%$ from $1\%$ (which also makes the attack more demanding), the improvement of attack effectiveness is less than $5\%$.
Considering both the attack effectiveness and the difficulty of conducting the attack, it is reasonable for us to set the default value of $\xi$ to be $1\%$.
From the general observation, we could conclude that FedRecAttack is effective with very small proportion of public interactions.
\begin{table}[h]
	\centering
	\caption{Impact of $\xi$ on Effectiveness of FedRecAttack.}
	\resizebox{0.495\textwidth}{!}{
		\begin{tabular}{|c|c|c|c|c|c|}
			\hline
			\multirow{2}[4]{*}{\textbf{Metric}} & \multicolumn{5}{c|}{\textbf{Proportion of Public Interactions}} \bigstrut\\
			\cline{2-6} & \multicolumn{1}{c|}{$\xi=1\%$} & \multicolumn{1}{c|}{$\xi=2\%$} &\multicolumn{1}{c|}{$\xi=3\%$} & \multicolumn{1}{c|}{$\xi=5\%$} & \multicolumn{1}{c|}{$\xi=10\%$} \bigstrut\\
			\hline 	 	 
			$\text{ER@5}$       & 0.9400 & 0.9818 & 0.9882 & 0.9936 & 0.9914\bigstrut\\
			\cline{1-6} 
			$\text{ER@10}$      & 0.9475 & 0.9893 & 0.9914 & 0.9946 & 0.9925\bigstrut\\
			\cline{1-6} 
			$\text{NDCG@10}$    & 0.9411 & 0.9789 & 0.9866 & 0.9886 & 0.9890\bigstrut\\
			\hline
		\end{tabular}
	}
	\label{table-impact-of-prior-knowledge}
\end{table}

\textbf{The proportion of malicious users ($\boldsymbol{\rho}$)} \quad
To explore the impact of the proportion of malicious users on the effectiveness of FedRecAttack, we set different values of $\rho$ in $\{1\%,2\%,3\%,5\%,10\%\}$.
It is evident from Table~\ref{table-impact-of-malicious-users} that the proportion of malicious users is a key factor to the effectiveness of FedRecAttack.
The effectiveness is improved significantly as the proportion increases.
Particularly, when $\rho=5\%$, $\text{ER@5}$ comes to a value of $0.9400$, which is just a little worse than the maximum.
Amongst other model poisoning attacks, $\rho=5\%$ is adequately small.
Remember that the attack cost is positively correlated with the proportion of malicious users.
Taking both aspects of effectiveness and cost of attack into consideration, we set the default value of $\rho$ to be $5\%$.
It could be concluded that FedRecAttack reaches quite good results with fairly small proportion of malicious clients. 
\begin{table}[h]
	\centering
	\caption{Impact of $\rho$ on Effectiveness of FedRecAttack.}
	\resizebox{0.495\textwidth}{!}{
		\begin{tabular}{|c|c|c|c|c|c|}
			\hline
			\multirow{2}[4]{*}{\textbf{Metric}} & \multicolumn{5}{c|}{\textbf{Proportion of Malicious Users}} \bigstrut\\
			\cline{2-6} 
			& \multicolumn{1}{c|}{$\rho=1\%$} & \multicolumn{1}{c|}{$\rho=2\%$} & \multicolumn{1}{c|}{$\rho=3\%$} & \multicolumn{1}{c|}{$\rho=5\%$} & \multicolumn{1}{c|}{$\rho=10\%$}\bigstrut\\
			\hline
			ER@5    & 0.0011  & 0.0043  & 0.6902  & 0.9400 & 0.9475\bigstrut\\
			\hline
			ER@10   & 0.0011  & 0.0075  & 0.7395  & 0.9475 & 0.9518\bigstrut\\
			\hline
			NDCG@10 & 0.0011  & 0.0042  & 0.6615  & 0.9411 & 0.9423\bigstrut\\
			\hline
		\end{tabular}
	}
	\label{table-impact-of-malicious-users}
\end{table}

\begin{table}[b]
	\centering
	\caption{Impact of $\kappa$ on Effectiveness of FedRecAttack.}
	\resizebox{0.495\textwidth}{!}{
		\begin{tabular}{|c|c|c|c|c|c|}
			\hline
			\multirow{2}[4]{*}{\textbf{Metric}} & \multicolumn{5}{c|}{\textbf{Maximum Number of Non-zero Rows}} \bigstrut\\
			\cline{2-6} & \multicolumn{1}{c|}{$\kappa=20$} & \multicolumn{1}{c|}{$\kappa=40$} & \multicolumn{1}{c|}{$\kappa=60$} & \multicolumn{1}{c|}{$\kappa=80$} & \multicolumn{1}{c|}{$\kappa=100$} \bigstrut\\
			\hline
			$\text{ER}@5$   & 0.9475  & 0.9464  & 0.9400  & 0.9507 & 0.9453 \bigstrut\\
			\hline
			$\text{ER}@10$  & 0.9539  & 0.9518  & 0.9475  & 0.9593 & 0.9518\bigstrut\\
			\hline
			$\text{NDCG@10}$& 0.9453  & 0.9442  & 0.9411  & 0.9480 & 0.9456 \bigstrut\\
			\hline
		\end{tabular}
	}
	\label{table-impact-of-non-zero-gradients}
\end{table}

\textbf{The maximum number of non-zero rows in uploaded poisoned gradients ($\boldsymbol{\kappa}$)}\quad 
We examine the results of different values of $\kappa$ in $\{20,40,60,80,100\}$.
As shown in Table~\ref{table-impact-of-non-zero-gradients}, $\kappa$ has little impact on attack effectiveness.
The validity of FedRecAttack can be proved by the fact that, for all values of $\kappa$, the attack effectiveness remains in high level (all metrics reach $94\%$).
For central server, more number of non-zero rows in user's uploaded gradients implies more number of items which the user has interacted with. 
If the number of non-zero rows in uploaded poisoned gradients is abnormally high, the attack is highly detectable.
On all three datasets, for most users, the number of interactions is around $30$, which indicates the number of non-zero rows in uploaded gradients is around $60$ (including both positive and negative item gradients).
Based on the above evidence, it is suitable to set $\kappa=60$ as default.

\subsection{Effectiveness of Attacks}
We compare the effectiveness of FedRecAttack with that of other baseline attacks in cases of different values of $\rho$ $\{3\%,5\%,10\%\}$ on three real-world datasets respectively.
Note that for FedRecAttack we set $\xi=1\%$.
As shown in Table~\ref{table-compare-to-baseline}, FedRecAttack achieves the best attack effectiveness on all three datasets with different proportions of malicious users.
In cases of small proportion of malicious users, other methods have merely slight effects, while FedRecAttack remains good performance.
From these evidences, we can draw the conclusion that FedRecAttack has the least dependency on the proportion of malicious users.
Moreover, from the table we can find that, the more dense the dataset is, the more difficult the attack is.
When the proportion of malicious users is kept constant, the attack effectiveness is higher on the more sparse dataset.

Additionally, we compare the effectiveness of FedRecAttack with that of two state-of-the-art data poisoning attacks on MovieLens-100K.
The two attacks are as following:
\begin{enumerate}
	\item P1~\cite{li2016data,fang2020influence}: data poisoning attack to matrix factorization based recommender systems.
	\item P2~\cite{huang2021data}: data poisoning attack to deep learning based recommender systems.
\end{enumerate}
For FedRecAttack, we keep the limitation on attacker's prior knowledge and set $\xi=1\%$.
For P1 and P2, because both of them highly rely on attacker's prior knowledge of user-item interactions and can not work with the limitation, we conduct the experiments with the same settings as in~\cite{huang2021data} (assuming attacker has access to all user-item interactions).
We set different values of $\rho$ in $\{0.5\%,1\%,3\%,5\%\}$ and adopt $\text{ER@10}$ to evaluate the attack effectiveness. 
As shown in Table~\ref{table-compare-on-ML100K}, P1 and P2 have better performance when the proportion of malicious users is extremely small.
We argue that they benifit from attacker's prior knowledge of all interactions.
However, $\text{ER@10}$ of these methods is hardly able to reach a satisfactory level, while that of FedRecAttack rapidly rises to a significant level as $\rho$ increases.
\begin{table}[b]
	\centering
	\caption{$\text{ER@10}$ of FedRecAttack and Other Data Poisoning Attacks on MovieLens-100K.}
	\resizebox{0.48\textwidth}{!}{
		\begin{tabular}{|c|c|c|c|c|}
			\hline
			\multirow{2}[4]{*}{\textbf{Attack Method}} & \multicolumn{4}{c|}{\textbf{Proportion of Malicious Users}} \bigstrut\\
			\cline{2-5}
			& $\rho=0.5\%$ & $\rho=1\%$ & $\rho=3\%$ & $\rho=5\%$ \bigstrut\\
			\hline
			None & 0.0000 & 0.0000 & 0.0000 & 0.0000 \bigstrut\\
			\hline
			P1 & 0.0001 & 0.0002 & 0.0014 & 0.0033 \bigstrut\\
			\hline
			P2 & 0.0007 & 0.0019 & 0.0111 & 0.0206 \bigstrut\\
			\hline
			\textbf{FedRecAttack} & \textbf{0.0000} & \textbf{0.0011} & \textbf{0.7449} & \textbf{0.9475} \bigstrut\\
			\hline
		\end{tabular}
	}
	\label{table-compare-on-ML100K}
\end{table}

\begin{table*}[t]
	\centering
	\caption{Effectiveness of Different Attacks with Different Proportions of Malicious Users.}
	\resizebox{0.99\textwidth}{!}{
		\begin{tabular}{|c|c|c|c|c|c|c|c|c|c|c|}
			\hline
			\multirow{3}[6]{*}{\textbf{Dataset}} & \multicolumn{1}{c|}{\multirow{3}[6]{*}{\textbf{Attack Method}}} & \multicolumn{9}{c|}{\textbf{Proportion of Malicious Users}} \bigstrut\\
			\cline{3-11} & & 
			\multicolumn{3}{c|}{$\rho=3\%$} & \multicolumn{3}{c|}{$\rho=5\%$} & \multicolumn{3}{c|}{$\rho=10\%$} \bigstrut\\
			\cline{3-11} & &
			\multicolumn{1}{p{1.2cm}<{\centering}|}{$\text{ER@5}$} & 
			\multicolumn{1}{p{1.2cm}<{\centering}|}{$\text{ER@10}$} & 
			\multicolumn{1}{p{1.2cm}<{\centering}|}{$\text{NDCG@10}$} & 
			\multicolumn{1}{p{1.2cm}<{\centering}|}{$\text{ER@5}$} & 
			\multicolumn{1}{p{1.2cm}<{\centering}|}{$\text{ER@10}$} & 
			\multicolumn{1}{p{1.2cm}<{\centering}|}{$\text{NDCG@10}$} & 
			\multicolumn{1}{p{1.2cm}<{\centering}|}{$\text{ER@5}$} &
			\multicolumn{1}{p{1.2cm}<{\centering}|}{$\text{ER@10}$} &
			\multicolumn{1}{p{1.2cm}<{\centering}|}{$\text{NDCG@10}$}
			\bigstrut\\
			\hline
			\multirow{5}[10]{*}{MovieLens-100K} 
			& None & 0.0000 & 0.0000 & 0.0000 & 0.0000 & 0.0000 & 0.0000 & 0.0000 & 0.0000 & 0.0000\bigstrut\\
			\cline{2-11}
			& Random & 0.0000 & 0.0000 & 0.0000 & 0.0000 & 0.0000 & 0.0000 & 0.0011 & 0.0011 & 0.0004\bigstrut\\
			\cline{2-11}
			& Bandwagon & 0.0011 & 0.0011 & 0.0011 & 0.0000 & 0.0021 & 0.0006& 0.0000 & 0.0000 & 0.0000 \bigstrut\\
			\cline{2-11}
			& Popular & 0.0011 & 0.0011 & 0.0005 & 0.0011 & 0.0011 & 0.0011 & 0.0032 & 0.0075 & 0.0035 \bigstrut\\
			\cline{2-11}
			& \textbf{FedRecAttack} & \textbf{0.6988} & \textbf{0.7449} & \textbf{0.6702} & \textbf{0.9400} & \textbf{0.9475} 
			& \textbf{0.9411} & \textbf{0.9507} & \textbf{0.9528} & \textbf{0.9455} \bigstrut\\
			\hline
			\multirow{5}[10]{*}{MovieLens-1M} 
			& None & 0.0000 & 0.0000 & 0.0000 & 0.0000 & 0.0000 & 0.0000 & 0.0000 & 0.0000 & 0.0000\bigstrut\\
			\cline{2-11}
			& Random & 0.0000 & 0.0000 & 0.0000 & 0.0002 & 0.0002 & 0.0001 & 0.0002 & 0.0005 & 0.0002 \bigstrut\\
			\cline{2-11}
			& Bandwagon & 0.0000 & 0.0000 & 0.0000 & 0.0000 & 0.0000 & 0.0000 & 0.0010 & 0.0012 & 0.0008 \bigstrut\\
			\cline{2-11}
			& Popular & 0.0035 & 0.0056 & 0.0030 & 0.0393 & 0.0503 & 0.0349 & 0.1358 & 0.1598 & 0.1255 \bigstrut\\
			\cline{2-11}
			& \textbf{FedRecAttack} & \textbf{0.9722} & \textbf{0.9752} & \textbf{0.9684} & \textbf{0.9659} & \textbf{0.9704} 
			& \textbf{0.9610} & \textbf{0.9689} & \textbf{0.9742} & \textbf{0.9646} \bigstrut\\
			\hline
			\multirow{5}[10]{*}{Steam-200K} 
			& None & 0.0000 & 0.0000 & 0.0000 & 0.0000 & 0.0000 & 0.0000 & 0.0000 & 0.0000 & 0.0000\bigstrut\\
			\cline{2-11} 
			& Random & 0.0027 & 0.0037 & 0.0022 & 0.0024 & 0.0029 & 0.0025 & 0.0029 & 0.0032 & 0.0027 \bigstrut\\
			\cline{2-11}
			& Bandwagon & 0.0133 & 0.0157 & 0.0121 & 0.0702 & 0.0952 & 0.0669 & 0.8829 & 0.8944 & 0.8774 \bigstrut\\
			\cline{2-11}
			& Popular & 0.2067 & 0.3129 & 0.1994 & 0.7165 & 0.7639 & 0.6908 & 0.8349 & 0.8480 & 0.8246 \bigstrut\\
			\cline{2-11}
			& \textbf{FedRecAttack} & \textbf{0.9843} & \textbf{0.9848} & \textbf{0.9833} & \textbf{0.9835} & \textbf{0.9848} 
			& \textbf{0.9831} & \textbf{0.9864} & \textbf{0.9869} &\textbf{0.9852} \bigstrut\\
			\hline
		\end{tabular}%
	}
	\label{table-compare-to-baseline}%
\end{table*}%

Moreover, we compare FedRecAttack with following state-of-the-art model poisoning attacks on MovieLens-1M:
\begin{enumerate}
	\item P3~\cite{bhagoji2019analyzing}: poison the federated learning model by directing it to misclassify the target input.
	\item P4~\cite{baruch2019little}: a genaral method for attacking distributed models with defense mechanisms.
	\item EB~\cite{zhang2021pipattack}: explicitly boost the predicted scores between malicious users and target items.
	\item PipAttack~\cite{zhang2021pipattack}: poison the federated recommender model with the information about items popularity.
\end{enumerate}
Since the above model poisoning attacks require more malicious users to be effective, we set $\rho=\{10\%,20\%,30\%,40\%\}$ respectively.
For FedRecAttack, we set $\xi=1\%$.
And for the others, we adopt the same settings as in~\cite{zhang2021pipattack} (assuming attacker has side information about items' popularity).
As presented in Table~\ref{table-compare-on-ML1M}, it is clear that $\text{ER@5}$ of P3, P4 and EB is numerically unstable with different proportions of malicious users, while both PipAttack and FedRecAttack maintain comparatively high effectiveness.
And when the proportion of malicious users is small (\textit{i.e.,} $\rho=10\%$), FedRecAttack has the best performance in all model poisoning attacks, which demonstrates the strengths of FedRecAttack.

\subsection{Stealthiness of Attacks}
\begin{figure}[t]
	\setlength{\abovecaptionskip}{-10pt}
	\centering
	\includegraphics[width=1\linewidth]{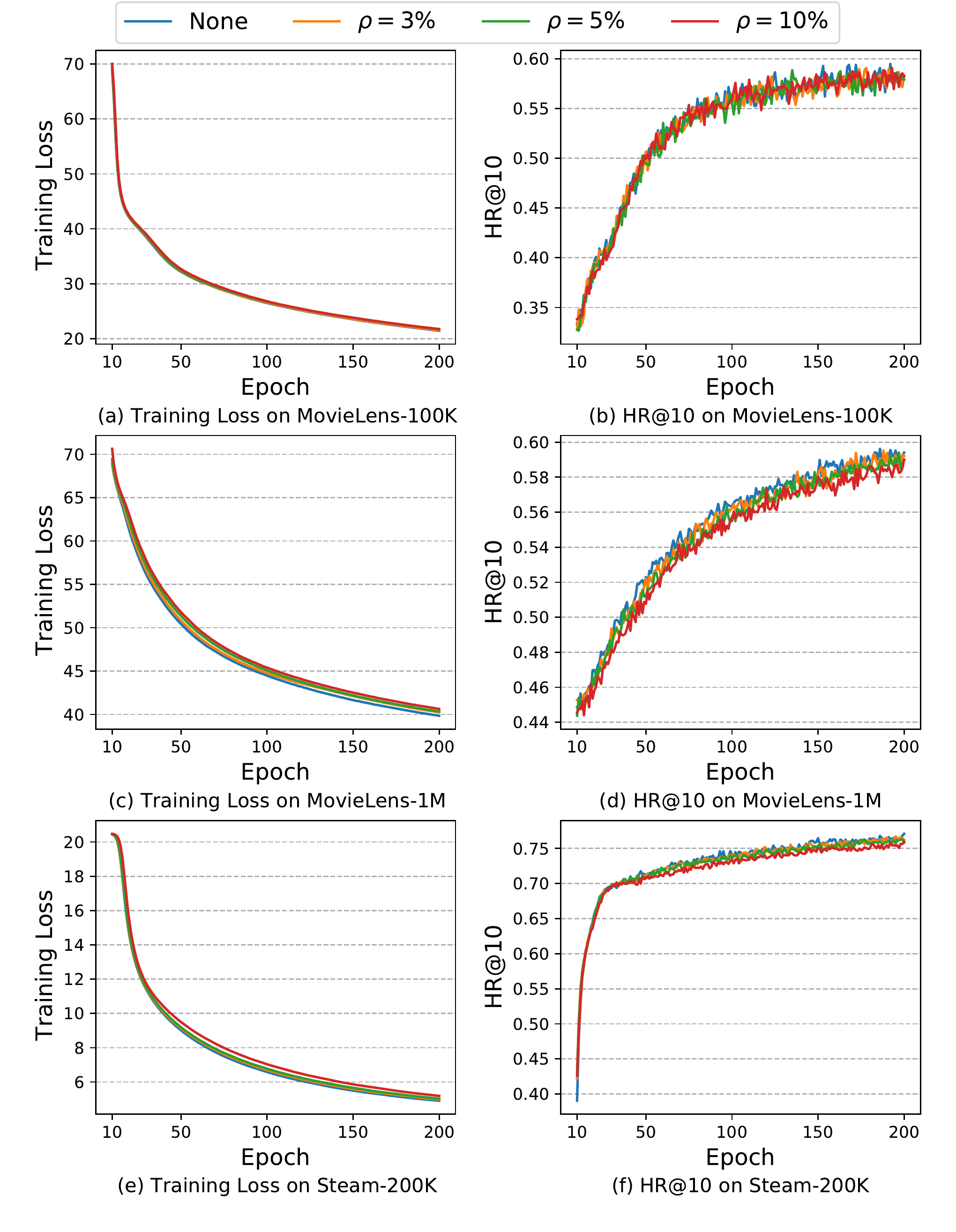}
	\caption{The side effects of FedRecAttack with different proportions of malicious users on MovieLens-100K, MovieLens-1M and Steam-200K.}
	\label{fig:performance-1M}
\end{figure}
\begin{table*}[t]
	\centering
	\caption{$\text{HR@10}$ and $\text{ER@5}$ of FedRecAttack and Other Model Poisoning Attacks on MovieLens-1M.}
	\resizebox{0.75\textwidth}{!}{
		\begin{tabular}{|c|c|c|c|c|c|c|c|c|}
			\hline
			\multirow{2}[10]{*}{\textbf{Attack Method}} 
			& \multicolumn{8}{c|}{\textbf{Proportion of Malicious Uers}}
			\bigstrut\\
			\cline{2-9}
			& \multicolumn{2}{c|}{$\rho=10\%$} & \multicolumn{2}{c|}{$\rho=20\%$} & \multicolumn{2}{c|}{$\rho=30\%$} & \multicolumn{2}{c|}{$\rho=40\%$} 
			\bigstrut\\
			\cline{2-9}
			& HR@10 & ER@5 & HR@10 & ER@5 & HR@10 & ER@5 & HR@10 & ER@5
			\bigstrut\\
			\hline
			None & 0.5940 & 0.0000 & 0.5940 & 0.0000 & 0.5940 & 0.0000 & 0.5940 & 0.0000\bigstrut\\
			\hline
			P3 & 0.4434 & 0.0000 & 0.4430 & 0.0000 & 0.4435 & 0.0154 & 0.4454 & 0.0298\bigstrut\\
			\hline
			P4 & 0.4392 & 0.0000 & 0.4386 & 0.9625 & 0.4320 & 0.9016 & 0.4425 & 1.0000\bigstrut\\
			\hline
			EB & 0.4432 & 0.0000 & 0.4449 & 1.0000 & 0.4363 & 0.9998 & 0.4432 & 1.0000\bigstrut\\
			\hline
			PipAttack & 0.4384 & 0.9513 & 0.4412 & 1.0000 & 0.4401 & 1.0000 & 0.4349 & 1.0000\bigstrut\\
			\hline
			\textbf{FedRecAttack} & \textbf{0.5901} & \textbf{0.9689} &\textbf{0.5800} & \textbf{0.9735} 
			& \textbf{0.5829} & \textbf{0.9733} & \textbf{0.5800} & \textbf{0.9786} \bigstrut\\
			\hline
		\end{tabular}
	}
	\label{table-compare-on-ML1M}
\end{table*}

Our work focuses not only on the effectiveness of attacks, but also on the stealthiness of attacks.
Methods to detect attacks in FR can be divided into two categories:
i) One line of the methods detects whether there are any abnormalities in the uploaded gradients.
ii) The other line of the methods observes whethere there is a suspicious degradation in recommendation accuracy during the training stage.
Unfortunately, the former line of methods does not perform well in FR because of the following two reasons:
i) The feature vectors of different users vary widely, hence the uploaded gradients vary widely.
ii) In Eq.~\eqref{Eq-add-noise}, as guided by the policy of differential privacy, user clients will add noise to their uploaded gradients.
Both of the two reasons increase the difficulty of distinguishing poisoned gradients from clean gradients.
In contrast, the latter line of methods is always much more practical in FR.
Different from data poisoning attacks, most model poisoning attacks work without attacker's prior knowledge of user-item interactions.
As a result, in model poisoning attacks, the poisoned gradients generated by attacker are not precise enough to raise $\text{ER@K}$ without any side effects on recommendation accuracy.
If there is a significant drop in recommendation accuracy, the attack would be detected easily.

Therefore, we emphasize the side effects of FedRecAttack on recommendation accuracy to ensure its stealthiness. 
We adopt $\text{HR@10}$, which is also used in~\cite{he2017neural}, as our metric to evaluate recommendation accuracy.
Fig.~\ref{fig:performance-1M} shows training loss and $\text{HR@10}$ in each training iteration under FedRecAttack and none attack.
In FedRecAttack, we set $\rho=\{3\%,5\%,10\%\}$ respectively.
As the figure shown, the side effects of FedRecAttack are negligible, which means FedRecAttack is stealthy.

Furthermore, we compare $\text{HR@10}$ under FedRecAttack with that under other model poisoning attacks on MovieLens-1M.
Since other model poisoning attacks require more malicious users to be effective, we set $\rho=\{10\%,20\%,30\%,40\%\}$ respectively.
As shown in Table \ref{table-compare-on-ML1M}, all other methods caused significant drops of $\text{HR@10}$ (more than $25\%$), which indicates these attacks are easy to notice. 
In contrast, FedRecAttack has the slightest side effects ($\text{HR@10}$ has only dropped less than $2.5\%$),
and its effectiveness always stays in high level (only a little worse than the best).
From these observations, we can conclude that FedRecAttack strikes a perfect balance between attack effectiveness and side effects.

We consider that, the function $g$ we used in Eq.~\eqref{eq-attack-loss-for-single-user} is the key that the side effects of FedRecAttack are slight.
As $x$ decreases, $g'(x)$ converges to $0$.
Due to this characteristic of $g$, in FedRecAttack, the predicted scores of target items will not be raised indefinitely, but will only be raised to exactly a little higher than that of the last item in user's recommendation list.

\begin{table}[h]
	\centering
	\caption{Effectiveness of FedRecAttack with \& without Attacker's Prior Knowledge of Public Interactions.}
		\begin{tabular}{|c|c|c|c|}
			\hline
			\multirow{2}[4]{*}{\textbf{Dataset}} & \multirow{2}[4]{*}{\textbf{Metric}}
			& \multicolumn{2}{c|}{\textbf{Proportion of Public Interactions}} \bigstrut\\
			\cline{3-4}
			& & \quad\quad$\xi=1\%$\quad\quad & \quad\quad$\xi=0\%$\quad\quad \bigstrut\\
			\hline
			\multirow{3}[4]{*}{MovieLens-100K}
			& $\text{ER@5}$ & 0.9400 & 0.0000 \bigstrut\\
			\cline{2-4}
			& $\text{ER@10}$ & 0.9475 & 0.0000 \bigstrut\\
			\cline{2-4}
			& $\text{NDCG@10}$ & 0.9411 & 0.0000 \bigstrut\\
			\hline
			\multirow{3}[4]{*}{MovieLens-1M}
			& $\text{ER@5}$ & 0.9659 & 0.0000 \bigstrut\\
			\cline{2-4}
			& $\text{ER@10}$ & 0.9704 & 0.0000 \bigstrut\\
			\cline{2-4}
			& $\text{NDCG@10}$ & 0.9610 & 0.0000 \bigstrut\\
			\hline
			\multirow{3}[4]{*}{Steam-200K}
			& $\text{ER@5}$ & 0.9835 & 0.0000 \bigstrut\\
			\cline{2-4}
			& $\text{ER@10}$ & 0.9848 & 0.0000 \bigstrut\\
			\cline{2-4}
			& $\text{NDCG@10}$ & 0.9831 & 0.0000 \bigstrut\\
			\hline
		\end{tabular}
	\label{table-ablation-test}
\end{table}

\subsection{Ablation Study}
In order to have a better understanding on the contributions of attacker's prior knowledge in FedRecAttack, we conduct an ablation experiment. 
We compare the effectiveness of FedRecAttack with public interactions ($\xi=1\%$) and without any ($\xi=0\%$) on all three datasets.
Although $\xi$ is very small, attacker's prior knowledge of public interactions is essential in FedRecAttack.
As shown in Table~\ref{table-ablation-test}, FedRecAttack is highly effective when $\xi=1\%$, while loses validity completely when $\xi=0\%$.
The reason of the above observation is that effective poisoned gradients can not be generated, as attacker can not rationally approximate users' feature matrix without attacker's prior knowledge.
From these evidences, we can testify our hypothesis of the validity of approximating users' feature matrix. 

\section{Conclusion and Future Work}
In this paper, we present FedRecAttack, a model poisoning attack against FR, which aims to make target items recommended to as many users as possible.
We evaluate the performance of FedRecAttack on three real-world datasets of different sizes.
The experimental results show that:
i) FedRecAttack achieves the state-of-the-art effectiveness.
ii) FedRecAttack has the slightest side effects among the existing model poisoning attacks.
iii) FedRecAttack still performs well when the proportion of malicious users and the proportion of public interactions are small.
From these observations, we can draw the conclusion that FedRecAttack can effectively poison the target recommender model at low cost with high stealthiness, demonstrating the vulnerability of FR.

The mainstream of existing methods to detect model poisoning attacks is training a support vector machine or a deep neural network to distinguish poisoned gradients from clean gradients~\cite{zhou2016svm}.
The mainstream of existing methods to defend against model poisoning attacks is adopting byzantine-robust aggregations (\textit{e.g.,} k-rum, trimmed mean, median)~\cite{yin2018byzantine}.
Although these methods have been studied in the field of federated learning, they do not fit FR perfectly.
In FR, because of the large differences in feature vectors of different users, the gradients uploaded by different users vary widely, which significantly increases the difficulty of detection or defense.
Interesting future work includes approaching methods of detecting or defending against model poisoning attacks with the consideration of the specificity of FR.

\section*{Acknowledgement}
This research is supported by the National Key R\&D Program of China (2021YFB2700500, 2021YFB2700501) and Scientific Research Fund of Zhejiang University (XY2021049).
The authors thank the anonymous reviewers for giving us a lot of valuable comments.

\bibliographystyle{IEEEtran}
\bibliography{main}

\end{document}